\documentclass[preprint,showpacs,preprintnumbers,amsmath,amssymb]{revtex4-1}

\usepackage{graphicx}
\usepackage{dcolumn}
\usepackage{bm}
\usepackage{booktabs}
\usepackage{amsmath,amssymb,mathrsfs}

\begin{document}


\def\preprint{1}

\title{Accelerating the convergence of free electron laser simulations by retrieving a spatially-coherent component of microbunching}

\author{Takashi Tanaka}\email{ztanaka@spring8.or.jp}

\affiliation{%
RIKEN SPring-8 Center, Koto 1-1-1, Sayo, Hyogo 679-5148, Japan
}%

\date{\today}

\begin{abstract}
A simple method to reduce the numerical cost in free electron laser (FEL) simulations is presented, which is based on retrieving a spatially-coherent component of microbunching to suppress artifact effects that can potentially overestimate the FEL gain; this significantly reduces the number of macroparticles to reach the numerical convergence and enables the direct computation of amplified radiation without solving the wave equation. Examples of FEL simulations performed to demonstrate the proposed method show that the computation time to get a reliable result is reduced by 1-2 orders of magnitude depending on the simulation condition.
\end{abstract}


\maketitle

\newcommand{\expe}{\mbox{e}^}
\newcommand{\dint}{\int\!\!\!\!\int}
\newcommand{\figw}{0.7}
\newcommand{\figh}{0.85}

\section{Introduction}
Numerical simulation is an important and indispensable tool to quantitatively evaluate the performances of free electron lasers (FELs), in which a set of equations describing the process of FEL amplification are numerically solved to quantify the formation of microbunching and amplification of radiation by the microbunched electron beam. These FEL equations fall into two groups; one describes the evolution of radiation copropagating with the electron beam, while the other gives the motion of an electron traveling along the undulator and interacting with the radiation.

The former equation, which is derived from the Maxwell's wave equation, has the form
\begin{equation}
\left[\bm{\nabla}_{\perp}^2+2i\frac{\omega}{c}\left(\frac{\partial}{\partial t}+\frac{1}{c}\frac{\partial}{\partial z}\right)\right]\bm{E}(\bm{r}_{\perp},z,t)=\bm{S}(\bm{r}_{\perp},z,t),
\label{eqn:wavedif}
\end{equation}
where $z$ and $\bm{r}_{\perp}=(x,y)$ are longitudinal and transverse coordinates, $t$ is time, $c$ is the speed of light, and $\omega$ is the angular frequency of radiation. The amplification of radiation is described by the growth of the complex amplitude $\bm{E}$, and the source term $\bm{S}$ drives the amplification process; the source term is defined by the spatial and temporal distributions of the electron beam together with its trajectory in the undulator, as will be discussed later in detail.

The latter equation gives the transverse position $(x_j, y_j)$, angle $(x'_j,y'_j)$, time $t_j$, and energy $\gamma_j$ of a specific ($j$-th) electron. Ideally, the equations are solved for all the electrons in the electron beam, which is usually too time consuming. Thus, instead of such a straightforward method, the electron beam is usually modeled by a collection of ``macroparticles,'' each of which represents a large number of electrons. In the simulation of self-amplified spontaneous emission (SASE) FELs, the macroparticles are initially distributed to correctly model the shotnoize in the electron beam (quiet loading) \cite{Fawley-PRAB-2002}.

To solve the wave equation (\ref{eqn:wavedif}) numerically, the source term $\bm{S}$ is discretized in the spatiotemporal coordinates; the electron beam is temporally divided into ``slices,'' each of which is further divided into ``cells'' in the transverse plane; the source term is then evaluated at the center of respective cells (grid points), by adding the contribution from each macroparticle over those contained inside the cell. Note that the length and interval of a slice are usually defined as a multiple of wavelength of radiation, so that the wave equation can be temporally averaged and modified to a convenient form to be numerically solved.

In the method to solve the wave equation using the macroparticle model explained above, we have two difficulties. First, the number of macroparticles should be sufficiently large to obtain a reliable simulation result, which potentially requires a long computation time. Second, the number of macroparticles needed for numerical convergence is sensitive to the simulation condition; for example, it grows rapidly as the FEL gain is reduced as shown later. 

To tackle the above issues, it is important to explore a scheme to accelerate the numerical convergence in FEL simulations. In this paper, we present a simple method for this purpose, which is based on retrieving a ``spatially-coherent'' component in the transverse microbunching profile to suppress artifact effects coming from an insufficient number of macroparticles. We also show that this method can be applied to reduce the numerical cost and computation time; the evolution of radiation can be easily described without rigorously solving the wave equation.

\section{Number of macroparticles for numerical convergence}\label{sec:example}
Let us first discuss the number of macroparticles to reach the numerical convergence in FEL simulations. As an example, we assume an electron beam whose temporal and spatial distribution functions are given by Gaussian functions, which drives a SASE FEL at the photon energy of 10 keV with the conditions summarized in Table 1. For reference, the number of electrons actually contained in the simulation time window of 18 fs is around $4\times 10^8$.

\begin{table}[h]
\begin{center}
\caption{Parameters assumed in the simulation examples.}
\vspace{5mm}
\begin{tabular}{lll}\hline \hline
Electron Beam Parameter &  \\ \hline
Electron Energy &  8 GeV \\
FWHM Bunch Length & 10 fs  \\ 
Bunch Charge & 0.065 nC  \\
Peak Current & 6 kA \\
Normalized Emittance & 0.4 $\mu$m$\cdot$rad  \\
Average Betatron Function & 12 m  \\
Average Beam Size & 17.5 $\mu$m \\
Energy Spread & $10^{-4}$ \\  \hline \hline
Undulator Parameter & \\ \hline
Period Length & 18 mm \\
Deflection Parameter &  2.18 \\
Segment Length & 5 m \\
Drift Length & 1.15 m \\
Number of Segments & 12 \\ \hline \hline
Simulation Condition &  \\ \hline
Photon Energy & 10 keV \\
Temporal Window & 18 fs \\
Spatial Window & 0.14$\times$0.14 mm$^2$  \\
Number of Grid Points & 65$\times$65 \\ \hline
\end{tabular}
\label{tbl:table1}
\end{center}
\end{table}

The FEL simulations have been performed with the computer code {\it SIMPLEX} \cite{Tanaka-JSR-2015} and {\it GENESIS} \cite{Reiche-NIMA-1999}; to evaluate the numerical convergence, simulations have been repeated with different numbers of macroparticles; the results are shown in Figs. 1(a) and 1(b), where the growth of the pulse energy along the undulator, or a gain curve, is plotted for each condition. The results with the two different simulation codes are consistent with each other, and are sensitive to the number of macroparticles in terms of the gain length and saturation power. It is reasonable to say that we need $10^7$ macroparticles to reach the numerical convergence in this condition.

It should be emphasized here that the numerical convergence discussed above is not universal, but is sensitive to a variety of factors. Among them, the harmonic number is one of the most sensitive. For example, Figs. 1(c) and 1(d) show the gain curves for the 3rd harmonic radiation simulated with the same conditions as the above; obviously, we need much more macroparticles than the fundamental radiation, and $10^8$ macroparticles or more are needed to reach the numerical convergence, which is close to the real electron number of  $4\times 10^8$.

\begin{figure}[ht]
\includegraphics[width=\figh\textwidth]{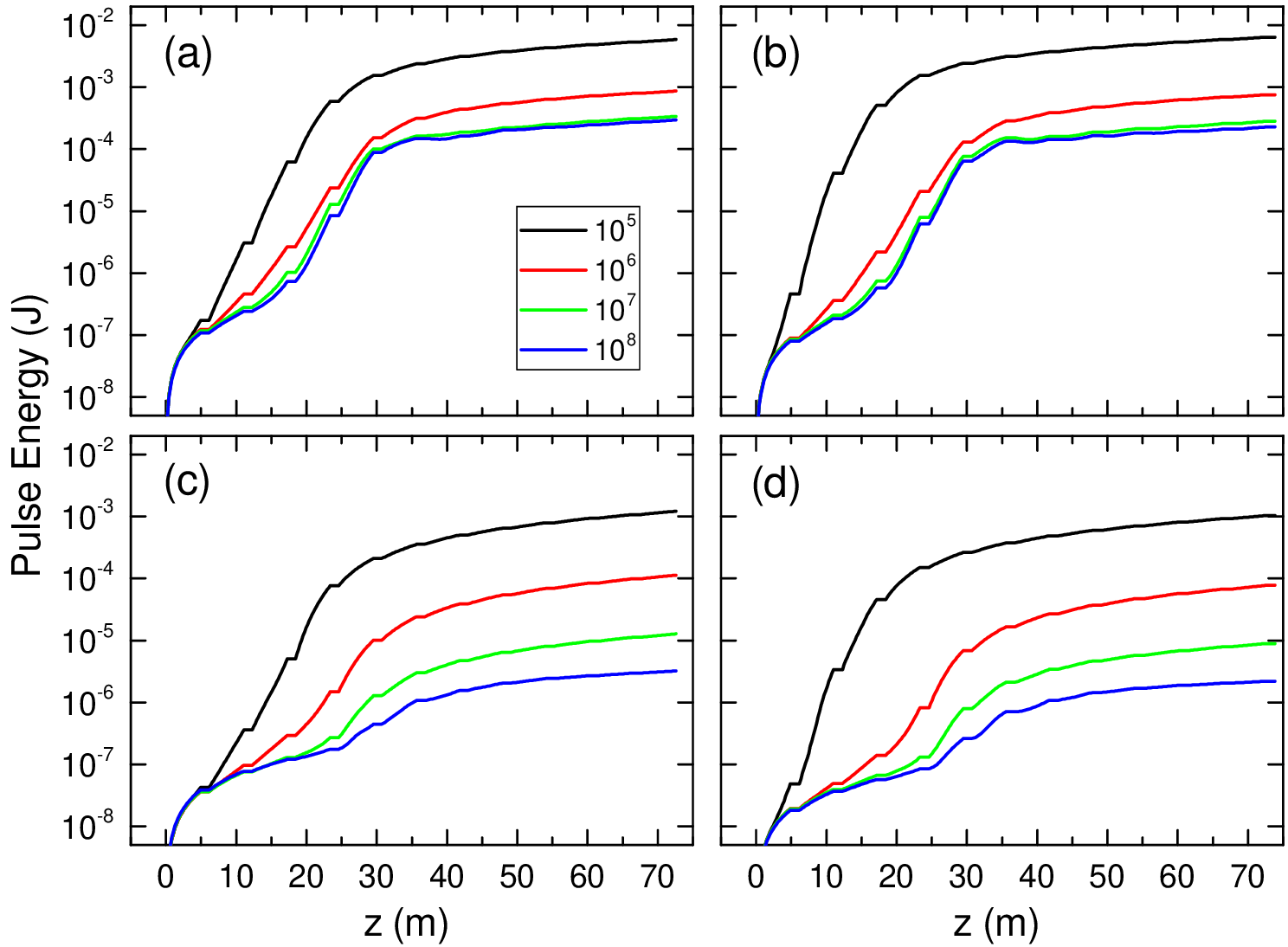}
\caption{FEL Gain curves simulated with (a,c) SIMPLEX and (b,d) GENESIS, using  different numbers of macroparticles for the (a,b) fundamental and (c,d) 3rd harmonic radiation.}
\end{figure}

It is worth noting that simulations with fewer macroparticles always overestimate the FEL gain; as discussed later in detail, this is attributable to artifacts in the source term $\bm{S}$ in equation (\ref{eqn:wavedif}), which comes from an insufficient number of macroparticles. In this paper, a numerical method is presented, which accelerates the numerical convergence by correctly evaluating $\bm{S}$ with much fewer macroparticles than what is usually required.

\section{Analytical formulation}
\subsection{Overestimation of the FEL gain by an insufficient number of macroparticles}
To investigate how the insufficient number of macroparticles results in an overestimation of the FEL gain, let us turn to the source term of the wave equation. According to a previous paper \cite{Tanaka-JSR-2015}, it has an explicit form
\begin{equation}
\bm{S}(\bm{r}_{\perp},z,t)=-iZ_0 I(z,t)\frac{\omega}{c}\left\langle\left(\bm{\beta}_{\bot j}+i\frac{c}{\omega}\nabla_{\bot}\right)\delta(\bm{r}_{\bot}-\bm{r}_{\bot j})\expe{-i\psi_j}\right\rangle_{z,t},
\label{eqn:src}
\end{equation}
where $Z_0$ is the vacuum impedance, $I$ denotes the beam current, $\bm{\beta}_{\bot j}$ and $\psi_j$ are the transverse velocity and phase of the $j$-th macroparticle. The bracket and subscripts $z$ and $t$ denote averaging over the macroparticles contained in a time range $[t-T/2,t+T/2]$ at the longitudinal position $z$, with $T$ being a temporal window for averaging. In the code {\it SIMPLEX}, the wave equations are solved by spatial Fourier transform and the radiation field is composed of two terms as
\begin{equation}
\bm{E}(\bm{r}_{\bot},z,s)=\bm{E}_d(\bm{r}_{\bot},z,s)+\bm{E}_g(\bm{r}_{\bot},z,s),
\end{equation}
where a new independent variable $s=z-\bar{v}_{z}t$ has been introduced instead of $t$, with $\bar{v}_{z}$ denoting the longitudinal velocity of a reference electron averaged over one undulator period. It is obvious that $s$ refers to the longitudinal position in the electron beam measured from the reference electron. 

The first term $\bm{E}_d$ describes the diffraction of radiation, and is given by
\begin{equation}
\bm{E}_d(\bm{r}_{\bot},z,s)=\mathscr{F}_s^{-1}\left\{\exp\left(-i\frac{c\bm{k}^2\Delta z}{2\omega}\right)\mathscr{F}_s\left[\bm{E}(\bm{r}_{\bot},z-\Delta z,s-\Delta s)\right]\right\},
\label{eqn:Ed}
\end{equation}
where $\Delta z$ is a step interval to solve the wave equation and $\Delta s=(1-\bar{v}_{z}/c)\Delta z$ corresponds to the slippage length of radiation while the electron travels the distance $\Delta z$ in the undulator. The operator $\mathscr{F}_s$ denotes the spatial Fourier transform with respect to $\bm{r}_{\bot}$, and $\bm{k}$ denotes the Fourier conjugate of $\bm{r}_{\bot}$. 

The second term $\bm{E}_g$ describes the amplification of radiation, and is given by
\begin{eqnarray}
\bm{E}_g(\bm{r}_{\bot},z,s)&=&-\frac{Z_0}{2}I(z-\Delta z,s-\Delta s) \nonumber \\
&\times&\mathscr{F}_s^{-1}\left\{\bm{F}\left(z,\frac{c\bm{k}}{\omega}\right)\mathscr{F}_s\left[\langle\delta(\bm{r}_{\bot}-\bm{r}_{\bot j})\expe{-i\psi_j}\rangle_{z-\Delta z,s-\Delta s}
\right]\right\},
\label{eqn:Eg0}
\end{eqnarray}
with
\begin{equation}
\bm{F}(z,\bm{\theta})=\int_{z-\Delta z}^{z}[\bm{\theta}-\bm{\beta}_{0 \perp}(z',z)]\expe{i\Phi(z',z)} dz',
\label{eqn:Fn}
\end{equation}
\begin{equation}
\Phi(z',z)=\frac{i\omega}{2c}\left[\int_0^{z'}\left\{\frac{1}{\gamma_0^2}+[\bm{\theta}-\bm{\beta}_{0 \perp}(z'')]^2\right\}dz''-\bm{\theta}^2 z\right],
\label{eqn:Fn}
\end{equation}
where $\gamma_0$ and $\bm{\beta}_{0 \perp}$ denote the Lorentz factor and transverse velocity of the reference electron.

Equation (\ref{eqn:Eg0}) means that the amplification of radiation is driven by two factors: $\bm{F}(z,\bm{\theta})$ and $\langle\delta(\bm{r}_{\bot}-\bm{r}_{\bot j})\expe{-i\psi_j}\rangle_{z,s}$. The former corresponds to the field amplitude of radiation, which is generated by a single electron while it travels from $z-\Delta z$ to $z$, and is emitted to the observation angle $\bm{\theta}$. The latter represents the microbunching at the transverse position $\bm{r}_{\bot}$ and slice position $s$.
 
For numerical operation of the above equations, we introduce a ``local bunch factor'' as follows
\begin{equation}
b(\bm{r}_{\bot},z,s)=\frac{1}{M\Delta x\Delta y}\int_{x-\Delta x/2}^{x+\Delta x/2}dx'\int_{y-\Delta y/2}^{y+\Delta y/2}dy'\sum_{|s_j-s|\leq\Delta s/2}^{M}\delta(\bm{r}_{\bot}'-\bm{r}_j)\expe{-i\psi_j},
\label{eqn:bh}
\end{equation}
where $\Delta x$ and $\Delta y$ are the horizontal and vertical intervals of the grid points, $M$ is the total number of macroparticles, and the summation is taken over macroparticles whose longitudinal coordinate $s_j$ satisfies the condition $|s_j-s|\leq \Delta s/2$. The local bunch factor obviously specifies the strength and phase of the microbunching averaged over a single cell, and we have a relation
\begin{equation}
\langle\delta(\bm{r}_{\bot}-\bm{r}_{\bot j})\expe{-i\psi_j}\rangle_{z,s}=b(\bm{r}_{\bot},z,s)\frac{M}{N(s)},
\end{equation}
where $N(s)$ is the number of macroparticles satisfying the condition $|s_j-s|\leq \Delta s/2$. Then, equation (\ref{eqn:Eg0}) reduces to
\begin{equation}
\bm{E}_g=-\frac{cqZ_0}{2\Delta s}\mathscr{F}_s^{-1}\left\{\bm{F}\left(z,\frac{c\bm{k}}{\omega}\right)\mathscr{F}_s\left[b(\bm{r}_{\bot},z-\Delta z,s-\Delta s)\right]\right\},
\label{eqn:Eg}
\end{equation}
where $q$ is the total charge contained in the simulation time window.

For later discussions, we introduce a slice bunch factor
\begin{equation}
B(z,s)=\int b(\bm{r}_{\bot},z,s)d\bm{r}_{\bot},
\end{equation}
to represent the microbunching in a target slice, and slice bunch power
\begin{equation}
\mathcal{P}(z,s)=\int |b(\bm{r}_{\bot},z,s)|^2d\bm{r}_{\bot},
\end{equation}
to evaluate the growth of the radiation power in that slice.

Now let us turn to how the microbunching evolves during amplification, and how it depends on the number of macroparticles. As an example, we consider bunch factors evaluated with $M=10^8$ and $M=10^6$ in the simulations presented in section \ref{sec:example}, at two different longitudinal positions of $z=5$ m (exit of the 1st segment) and $z=23.45$ m (4th segment). To facilitate the following discussions, let (a) specify the condition of $z=5$ m and $M=10^8$, (b) $z=5$ m and $M=10^6$, (c) $z=23.45$ m and $M=10^8$, and (d) $z=23.45$ m and $M=10^6$.

\begin{figure}[htb]
\includegraphics[width=\figw\textwidth]{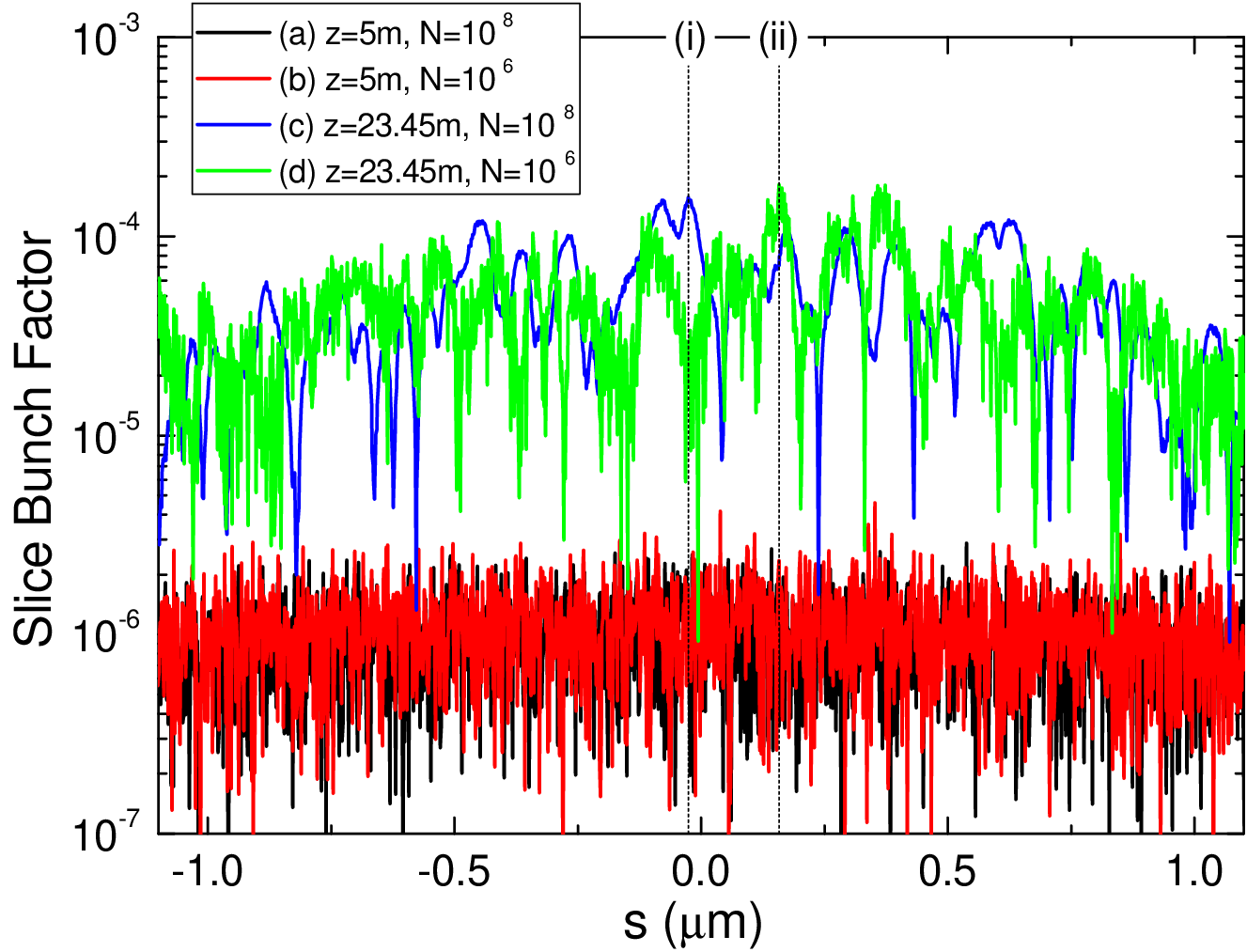}
\caption{Temporal profiles of $|B|$ at $z=5$ m with $M=10^8$ (black) and $M=10^6$ (red), and $z=23.45$ m with $M=10^8$ (blue) and $M=10^6$ (green).}
\end{figure}

Figure 2 shows the temporal profiles of the slice bunch factor $|B|$ in the four different conditions (a)-(d); we find that $|B|$ is enhanced by nearly two orders of magnitude after the electron beam passes through 3 undulator segments (from 2nd to 4th). Except for the spikes intrinsic to the SASE process, $M$ does not have a large impact on the maximum value and temporal average of $|B|$; this suggests that $M=10^6$ is sufficiently large as far as the slice bunch factor $B$ is concerned.

To clarify the reason why $M=10^6$ overestimates the FEL gain, let us investigate how the spatial profile of the bunch factor varies as the simulation conditions. For comparison between different conditions, we introduce a normalized local bunch factor
\begin{equation}
\hat{b}(\bm{r}_{\bot},z,s)=\frac{|B(z,s)|}{B(z,s)}b(\bm{r}_{\bot},z,s),
\end{equation}
so that 
\begin{equation}
\dint \hat{b}(\bm{r}_{\bot},z,s)d\bm{r}_{\bot}=|B(z,s)|,
\end{equation}
meaning that the spatial integral of the real part of $\hat{b}$ is always positive, while the imaginary part vanishes. Thus, we investigate the real part of $\hat{b}$, because it gives the information on the slice bunch factor as well as the local one. In the following discussion, the real part of $\hat{b}$ is simply referred to as $\hat{b}$ for simplicity.

Let us consider a target slice position to investigate the spatial profile of the bunch factor. It is reasonable to focus on the slice position where $|B|$ reaches the maximum at $z=23.45$ m, which is indicated by the dashed line in Fig. 2 for each value of $M$: (i) $s=-0.025$ $\mu$m for (a) and (c) ($M=10^8$), and (ii) $s=0.16$ $\mu$m for (b) and (d) ($M=10^6$).  

\begin{figure}[b]
\includegraphics[width=\textwidth]{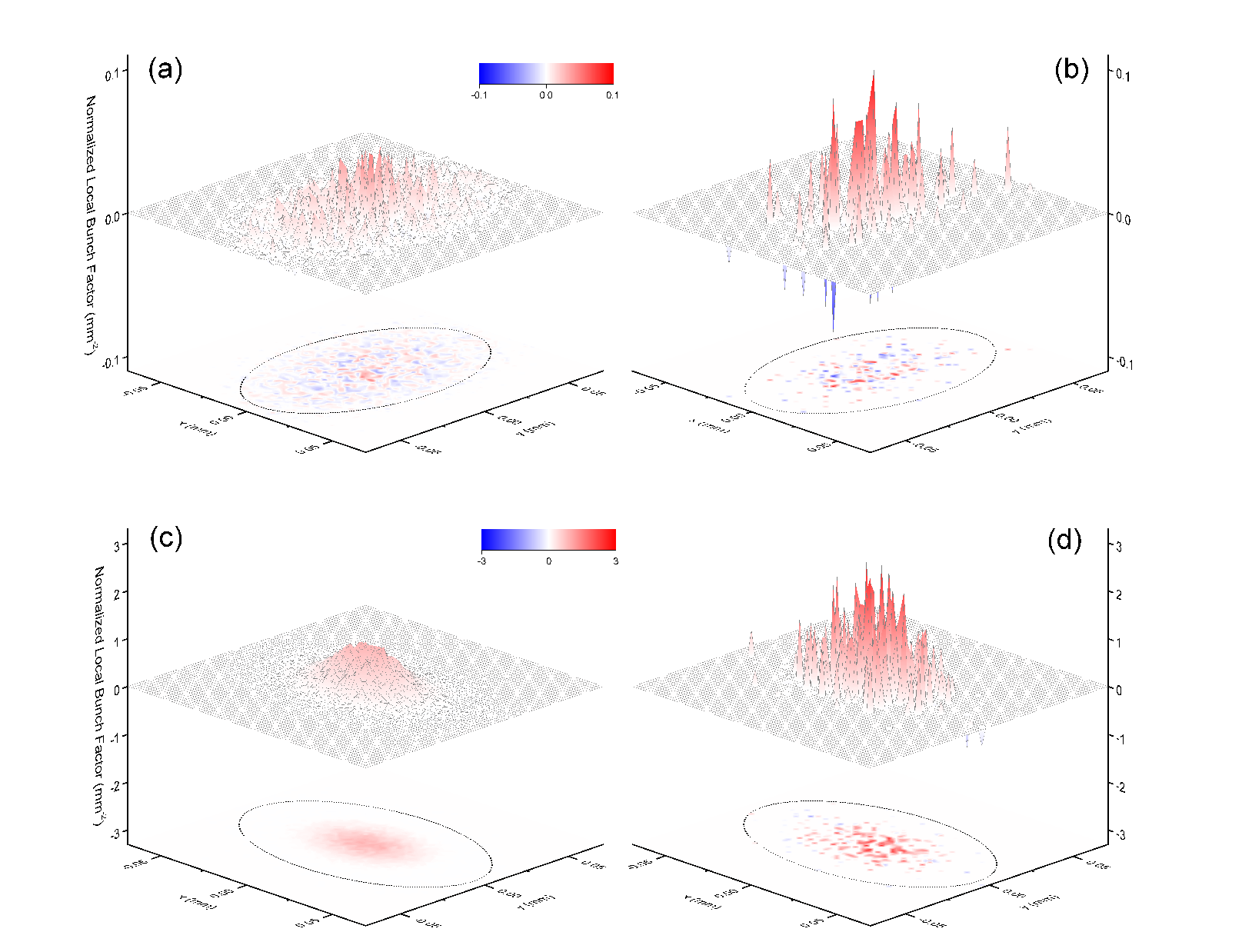}
\caption{Spatial profiles of the normalized local bunch factor retrieved from the simulation results with (a) $M=10^8$ and (b) $M=10^6$ at $z=5$ m; (c,d) same as (a,b) but at $z=23.45$ m.}
\end{figure}

Figures 3(a)-3(d) show the spatial profiles of the normalized local bunch factor $\hat{b}$ for the four different conditions (a)-(d), respectively. Note that $\hat{b}$ is averaged over 5 slices around the target slice position defined above, which corresponds to the coherence length \cite{Bonifacio-PRL-1994} (8 nm) in the condition under discussion; this is to improve the visualization without modifying the characteristics of $\hat{b}$. The ellipse in the contour plot indicates the area defined by $\pm 3\sigma_x$ and $\pm 3\sigma_y$, where $\sigma_x$ and $\sigma_y$ are the root mean square (RMS) electron beam sizes in the horizontal and vertical directions, with $\sigma_x=0.014$ (0.022) mm and $\sigma_y=0.022$ (0.014) at $z=5$ (23.45) m.

As found in Fig. 3(a), $\hat{b}$ at $z=5$ m with $M=10^8$ oscillates rapidly and randomly within an envelope defined by the distribution function of the electron beam, or a 2-dimensional (2D) Gaussian function with the RMS widths of $\sigma_x$ and $\sigma_y$. This means that the bunch factor at the exit of the 1st undulator segment is dominated by the shotnoize of the electron beam. The above discussion also applies to the simulation result with $M=10^6$ as found in Fig. 3(b); however, the maximum value is considerably larger than that with $M=10^8$. Recalling that the slice bunch factor $|B|$ is similar in each condition, it is obvious thus this difference comes from the fewer macroparticles contained in a single cell.

Figure 3(c) is the same as 3(a), but at $z=23.45$ m; through the amplification process, the maximum value of $|b|$ is enhanced by a factor of $\sim$30 compared with that at $z=5$ m. What is more important is that the spiky structure in the early stage of amplification found in Fig. 3(a) almost disappears, and roughly speaking, $\hat{b}$ is overall positive in the whole transverse plane. This means that the longitudinal positions ($s$) of the macroparticles contained in the same slice are close to each other, regardless of their transverse positions ($\bm{r}_{\bot}$). Thus, the microbunching in this condition is spatially in phase, or its ``spatial coherence'' is high.

Figure 3(d) is the same as 3(b), but with $M=10^6$. Although it reproduces the result with $M=10^8$ in terms of the constant phase over the transverse plane and enhancement factor of $\sim$30, the spiky structure still remains. This obviously comes from an insufficient number of macroparticles; $\hat{b}$ vanishes in cells where no macroparticles are contained, while it can be artificially higher than what is really reached in those with a reasonable number of macroparticles. Because of such a spiky structure of $\hat{b}$, the slice bunch power $\mathcal{P}$ evaluated with $M=10^6$ is about 3 times higher than that with $M=10^8$, although the slice bunch factor $|B|$ is similar; as a result, the simulation with $M=10^6$ overestimates the radiation power and FEL gain.
 
\subsection{Retrieving a spatially-coherent component of microbunching to suppress artifact effects}\label{sec:formula1}
To suppress the overestimation of the FEL gain due to an insufficient number of macroparticles, we need to retrieve the ``true'' spatial profile of the local bunch factor, as found in Fig. 3(c). For this purpose, let us assume that $b$ can be decomposed into two terms,
\begin{equation}
b(\bm{r}_{\bot},s)=b_{C}(\bm{r}_{\bot},s)+b_{S}(\bm{r}_{\bot},s),
\end{equation}
where $b_{C}$ denotes the spatially-coherent component and is a slowly varying function of $\bm{r}_{\bot}$, while $b_{S}$ denotes the incoherent one and rapidly oscillates as $\bm{r}_{\bot}$; note that the argument $z$ has been omitted because it is fixed in the following discussions. When integrated over $\bm{r}_{\bot}$, $b_{C}$ reduces to $B$, while $b_{S}$ vanishes. Obviously, the incoherent term $b_{S}$ has no impact on the amplification process because radiation coming from this component rapidly diverges because of diffraction, and thus can be omitted from equation (\ref{eqn:Eg}) to describe the evolution of radiation. In addition, it is natural to assume
\begin{equation}
\dint|b(\bm{r}_{\bot},s)|^2d\bm{r}_{\bot}=\dint|b_{C}(\bm{r}_{\bot},s)|^2d\bm{r}_{\bot}+\dint|b_{S}(\bm{r}_{\bot},s)|^2d\bm{r}_{\bot},
\label{eqn:bhs1}
\end{equation}
meaning that the cross term $b_{C}b_{S}^*$ vanishes when integrating over $\bm{r}_{\bot}$.

Now let us assume that $b_{C}$ is given by a simple 2D Gaussian function
\begin{equation}
b_{C}(\bm{r}_{\bot},s)=\frac{B(s)}{2\pi\kappa^2\sigma_x\sigma_y}\exp\left(-\frac{x^2}{2\kappa^2\sigma_x^2}-\frac{y^2}{2\kappa^2\sigma_y^2}\right),
\label{eqn:Gauss}
\end{equation}
where $\kappa$ is a scaling factor to define the horizontal and vertical RMS sizes of the 2D Gaussian function. The validity of this assumption is evident from Figs. 3(c); except for the spikes, $b$ is obviously represented by a 2D Gaussian function, whose aspect ratio is the same as that of the spatial distribution function of the electron beam.

To evaluate the scaling factor $\kappa$, we expand $b$ by Laguerre-Gaussian (LG) functions as follows
\begin{equation}
b(\bm{r}_{\bot},s)=\sum_{q}\frac{g_{q}(s)}{\sigma_x\sigma_y}\Psi_q\left(\frac{x}{\sigma_x},\frac{y}{\sigma_y}\right),
\end{equation}
with
\begin{equation}
\Psi_q(u,v)=\frac{1}{\sqrt{\pi}}L_q(u^2+v^2)\exp\left(-\frac{u^2}{2}-\frac{v^2}{2}\right),
\end{equation}
where $L_q$ is a Laguerre function of the $q$-th order.  Recalling the orthonormality of $\Psi_q$, the coefficient $g_{q}$ is given by
\begin{equation}
g_{q}(s)=\dint b(\bm{r}_{\bot},s)\Psi_q\left(\frac{x}{\sigma_x},\frac{y}{\sigma_y}\right)dxdy=\frac{1}{M}\sum_{|s_j-s|\leq\Delta s/2}^{M}\Psi_q\left(\frac{x_j}{\sigma_x},\frac{y_j}{\sigma_y}\right)\expe{-i\psi_j},
\label{eqn:gps}
\end{equation}
where equation (\ref{eqn:bh}) has been used. Now we have
\begin{equation}
\dint|b(\bm{r}_{\bot},s)|^2d\bm{r}_{\bot}=\frac{1}{\sigma_x\sigma_y}\sum_{q}|g_{q}(s)|^2.
\label{eqn:bhs2}
\end{equation}
Comparing equations (\ref{eqn:bhs1}) and (\ref{eqn:bhs2}), we can define a maximum order $Q$ of the LG function comprising the coherent term, namely
\begin{equation}
\dint|b(\bm{r}_{\bot},s)_C|^2d\bm{r}_{\bot}=\frac{1}{\sigma_x\sigma_y}\sum_{q=0}^Q|g_{q}(s)|^2,
\end{equation}
while the incoherent term is given by
\begin{equation}
\dint|b(\bm{r}_{\bot},s)_S|^2d\bm{r}_{\bot}=\frac{1}{\sigma_x\sigma_y}\sum_{q=Q+1}^{\infty}|g_{q}(s)|^2.
\end{equation}
Under the assumption (\ref{eqn:Gauss}) and averaging over the whole slices, we have
\begin{equation}
\kappa^2=\frac{\displaystyle \int |B(s)|^2ds}{4\pi \displaystyle \int\sum_{q=0}^Q |g_{p}(s)|^2 ds},
\end{equation}
where $Q$ should be determined to be consistent with the assumption that  $b_{C}$ is a slowly varying function of $\bm{r}_{\bot}$; to be more specific, $Q$ should be reasonably low, because $b_{C}$ approaches to $b$ for $q\rightarrow\infty$. 

To look for the reasonable value for $Q$, let us consider a simple case when $b_{S}(\bm{r}_{\bot},s)=0$ and thus the coefficient $g_{q}$ is given by
\begin{equation}
g_{q}(s)\propto\dint \exp\left(-\frac{u^2+v^2}{2\kappa^2}\right)\Psi_q(u,v)dudv,
\end{equation}
and examine how $g_{q}$ varies as the order $q$ and scaling factor $\kappa$. Figure 4 shows $|g_q|^2$ numerically evaluated for different values of $q$ in a practical range of $0.5\leq \kappa \leq 1$, normalized by the 0-th order. Because of the orthogonality of the LG function, $g_q$ vanishes for $q \geq 1$ and $\kappa=1$, and not shown explicitly.

\begin{figure}[htb]
\includegraphics[width=\figw\textwidth]{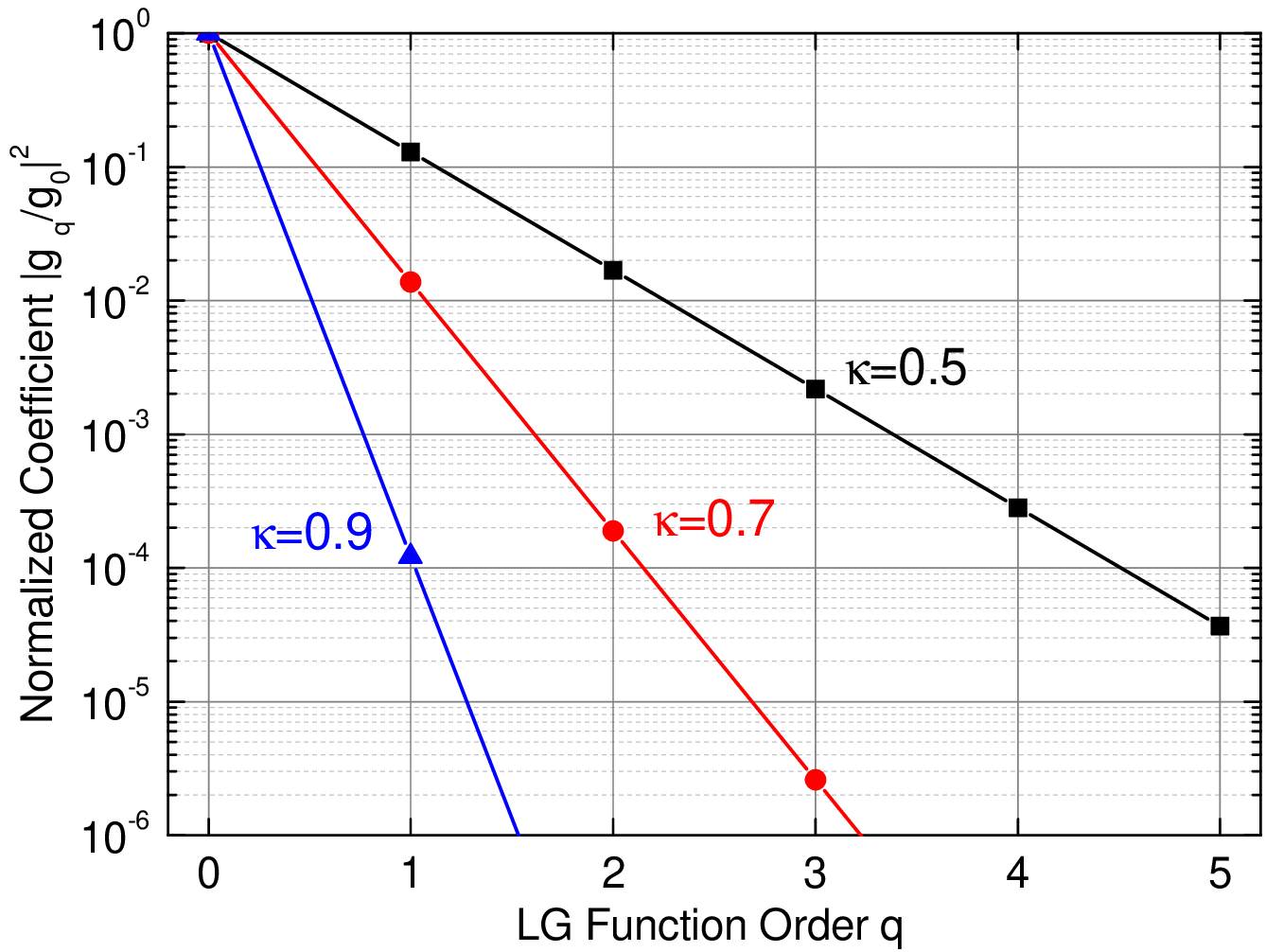}
\caption{Square of the coefficient $g_q$ for the $q$-th order LG function to expand a simple 2D Gaussian function for different scaling factors $\kappa$.}
\end{figure}

The numerical results show that $|g_q|^2$ drops rapidly as $q$, and the contribution from the 2nd and higher orders is negligibly small; even with an extreme  case when the transverse size of the local bunch factor is half of that of the electron beam ($\kappa=0.5$, black),  $|g_2|^2$ is two orders of magnitude lower than the fundamental component. Then we can conclude that $Q=1$ is the reasonable choice to define $\kappa$ and thus we finally have
\begin{equation}
\kappa^2=\frac{\displaystyle \int |B(s)|^2ds}{4\pi \displaystyle \int[|g_{0}(s)|^2+|g_{1}(s)|^2] ds}.
\label{eqn:kappa}
\end{equation}

Now, the spatially-coherent component can be retrieved from equations (\ref{eqn:Gauss}), (\ref{eqn:gps}) and (\ref{eqn:kappa}), using the distribution of the macroparticles. Then we have
\begin{equation}
\mathscr{F}_s[b(\bm{r}_{\bot},s)]=B(s)\exp\left(-\frac{X^2\theta_x^2+Y^2\theta_y^2}{2}\right),
\label{eqn:bfft}
\end{equation}
with $X=\kappa k\sigma_x$ and $Y=\kappa k\sigma_y$ denoting the normalized RMS widths of the local bunch factor in the horizontal and vertical directions. This formula can be used in (\ref{eqn:Eg}) to describe the amplification of radiation. Because it does not contain any artifacts coming from an insufficient number of macroparticles, it is expected that the overestimation of the FEL gain is avoided and the numerical convergence is reached with much fewer macroparticles.

\subsection{Direct calculation of the radiation field without solving the wave equation}\label{sec:formula2}
Besides accelerating the numerical convergence, the numerical method presented in the previous section can be applied to reduce the computation time by taking advantage of analytical formulas instead of performing a number of numerical processes.

Using equations (\ref{eqn:Eg}) and (\ref{eqn:bfft}), we have
\begin{equation}
\mathscr{F}_s[\bm{E}_g(\bm{r}_{\bot},z_n,s_m)]=-\frac{cqZ_0}{2\Delta s}B(z_{n-1},s_{m-1})\bm{F}(z_n,\bm{\theta})G_n(\bm{\theta}),
\label{eqn:angeg}
\end{equation}
with
\begin{equation}
G_n(\bm{\theta})=\exp\left(-\frac{X_n^2\theta_x^2+Y_n^2\theta_y^2}{2}\right),
\end{equation}
where subscripts $n$ in $X$ and $Y$ indicate that they should be evaluated at the longitudinal position $z=z_n$. Equation (\ref{eqn:angeg}) describes the amplification of radiation at the $m$-th slice and $n$-th step in the angular domain. 

To simplify the above formula, we compare the two functions $\bm{F}(z_n,\bm{\theta})$ and $G_n(\bm{\theta})$ in terms of the angular divergence. It is easy to understand that the typical divergences of $\bm{F}$ and  $G_n$ are $\gamma_0^{-1}$ and $X_n^{-1}$ (or $Y_n^{-1}$); the former is much larger than the latter as long as the electron beam size satisfies the condition $\sigma_{x,y}\gg \gamma_0\lambda/(2\pi\kappa)$ with $\lambda$ being the wavelength of radiation, which is usually the case for an electron beam in FELs. Then it is reasonable to make an approximation 
\begin{equation}
\bm{F}(z_n,\bm{\theta})G_n(\bm{\theta})\rightarrow\bm{F}(z_n,\bm{0})G_n(\bm{\theta}),
\label{eqn:approxF}
\end{equation}
and we have
\begin{equation}
\mathscr{F}_s[\bm{E}_g(\bm{r}_{\bot},z_n,s_m)]=\bm{e}_{n,m}G_n(\bm{\theta}),
\end{equation}
with
\begin{equation}
\bm{e}_{n,m}=-\frac{cqZ_0}{2\Delta s}B(z_{n-1},s_{m-1})\bm{F}(z_n,0),
\end{equation}
denoting the on-axis ($\bm{\theta}=\bm{0}$) radiation field in the angular domain, which is defined once the slice bunch factor is obtained. 

The angular representation $\bm{\varepsilon}_{p,m}(\bm{\theta})\equiv\mathscr{F}_s[\bm{E}(\bm{r}_{\bot},z_p,s_m)]$ at the $m$-th slice and $p$-th step is evaluated by summing up the contributions from the 1st to $p$-th steps. Recalling the slippage and diffraction effects, we have
\begin{equation}
\bm{\varepsilon}_{p,m}(\bm{\theta})=\sum_{n=1}^{p}\bm{e}_{n,m-p+n}G_n(\bm{\theta})\exp\left(-\frac{i\hat{z}_{pn}\bm{\theta}^2}{2}\right),
\label{eqn:ange}
\end{equation}
with $\hat{z}_{pn}=k(z_p-z_n)$ denoting the normalized distance. Because of its manageable form, the above formula significantly reduces the numerical cost to describe the amplified radiation. For example, the instantaneous radiation power $P_{p,m}$ given by
\begin{equation}
P_{p,m}=\frac{2}{Z_0\lambda^2}\dint |\bm{\varepsilon}_{p,m}(\bm{\theta})|^2d\bm{\theta},
\end{equation}
contains a spatial integral that can be analytical performed as follows
\begin{equation}
\dint G_i(\bm{\theta})G_ j(\bm{\theta})\exp\left(-\frac{i\hat{z}_{ij}\bm{\theta}^2}{2}\right)d\bm{\theta}=\frac{2\pi}{\sqrt{X_i^2+X_j^2+i\hat{z}_{ij}}\sqrt{Y_i^2+Y_j^2+i\hat{z}_{ij}}},
\end{equation}
meaning that the instantaneous radiation power is evaluated by summing up a number of terms instead of the 2-dimensional integral. Another example is the radiation field needed to compute the energy variation of each macroparticle interacting with radiation; inverse Fourier transform of (\ref{eqn:ange}) yields the radiation field $\bm{E}(\bm{r}_{\bot},z_p,s_m)$ at the $m$-th slice and $p$-th step, which can be analytically performed to give
\begin{eqnarray}
\bm{E}(\bm{r}_{\bot},z_p,s_m)=\sum_{n=1}^{p}
\frac{k^2\bm{e}_{n,m-p+n}}{2\pi\sqrt{X_n^2+i\hat{z}_{pn}}\sqrt{Y_n^2+i\hat{z}_{pn}}}\exp\left(-\frac{k^2x^2/2}{X_n^2+i\hat{z}_{pn}}-\frac{k^2y^2/2}{Y_n^2+i\hat{z}_{pn}}\right),
\end{eqnarray}
meaning that we do not have to solve the wave equations (\ref{eqn:Ed}) and (\ref{eqn:Eg0}). 

\subsection{Extension to high harmonics}
Although the discussions so far are focused on the fundamental radiation ($\lambda=\lambda_1$, where $\lambda_1$ is the fundamental wavelength of radiation), they can be applied to high-harmonic radiation by modifications $\lambda\rightarrow \lambda_1/h$ and $\psi\rightarrow h\psi$, with $h$ being a harmonic number. It should be noted, however, that the approximation (\ref{eqn:approxF}) is not valid for even-harmonic radiation that is canceled out in the forward direction $\bm{\theta}=\bm{0}$, and we have $\bm{F}(z,\bm{0})=\bm{0}$ and $\bm{\varepsilon}_{p,m}(\bm{\theta})=\bm{0}$, meaning that the amplification of even harmonics cannot be described correctly.

To overcome the above difficulty, we make an approximation
\begin{equation}
\bm{F}(z,\bm{\theta})=\left[\frac{\partial \bm{F}(z,\bm{\theta})}{\partial \theta_x}\right]_{\theta=\bm{0}}\theta_x+\left[\frac{\partial \bm{F}(z,\bm{\theta})}{\partial \theta_y}\right]_{\theta=\bm{0}}\theta_y,
\end{equation}
for even harmonics, to yield
\begin{equation}
\bm{\varepsilon}_{p,m}(\bm{\theta})=\sum_{n=1}^{p}(\bm{e}_{x;n,m-p+n}\theta_x+\bm{e}_{y;n,m-p+n}\theta_y)G_n(\bm{\theta})\exp\left(-\frac{i\hat{z}_{pn}\bm{\theta}^2}{2}\right),
\label{eqn:angeh}
\end{equation}
with
\begin{equation}
\bm{e}_{x;n,m}=-\frac{cqZ_0}{2\Delta s}B(z_{n-1},s_{m-1})\left[\frac{\partial \bm{F}(z,\bm{\theta})}{\partial \theta_x}\right]_{\theta=\bm{0}},
\end{equation}
and a similar expression for $\bm{e}_{y;n,m}$. The above formula makes it possible to derive analytical formulas for the radiation field and power, $\bm{E}$ and $P_{p,m}$. For example, $P_{p,m}$ contains a spatial integral that can be analytically performed as follows
\begin{equation}
\dint \theta_x^2G_i(\bm{\theta})G_ j(\bm{\theta})\exp\left(-\frac{i\hat{z}_{ij}\bm{\theta}^2}{2}\right)d\bm{\theta}=\frac{2\pi}{(X_i^2+X_j^2+i\hat{z}_{ij})^{3/2}\sqrt{Y_i^2+Y_j^2+i\hat{z}_{ij}}},
\end{equation}
and thus the radiation power of even-harmonic radiation is evaluated in the same manner as the fundamental radiation. Inverse Fourier transform of (\ref{eqn:angeh}) can be analytically performed to evaluate $\bm{E}(\bm{r}_{\bot},z_p,s_m)$ as well; in practice, however, this process can be skipped, because the radiation power of even-harmonic radiation is usually about 3 orders of magnitude lower than that of the fundamental radiation, and thus has negligible impact on the energy variation of each macroparticle.

\subsection{Contribution from the incoherent component}
The numerical methods presented in the former sections help to accelerate the numerical convergence and reduce the computation time. It should be noted, however, that the contribution from the incoherent component is neglected. Although this is valid in the description of the amplification of radiation, the incoherent component should be taken into account to evaluate the total radiation power in SASE FELs precisely, because it can be dominant in the early stage of amplification. To be more specific, radiation power coming from the incoherent component should be evaluated independently and added to that from the coherent component.

First, radiation field $\bm{E}_{S}$ generated by a single electron moving from $z=z_{n-1}$ to $z=z_n$ is computed using equation (\ref{eqn:Eg}), by substituting $\mathscr{F}_s[b(\bm{r}_{\bot})]=1$ and $I=ec/\Delta s$. Namely, we have 
\begin{equation}
\bm{E}_{S}=-\frac{ceZ_0}{2\Delta s}
\mathscr{F}_s^{-1}\left\{\bm{F}\left(z_n,\frac{c\bm{k}}{\omega}\right)\right\}.
\end{equation}
The corresponding radiation power is given as
\begin{equation}
P_S=\frac{2}{Z_0}\dint |\bm{E}_{S}|^2 d\bm{r}_{\bot}=\frac{Z_0}{2}\left(\frac{ce}{\lambda\Delta s}\right)^2\dint |\bm{F}(z_n,\bm{\theta})|^2d\bm{\theta}.
\end{equation}
Then the radiation power brought by $N_e$ electrons is given by $N_eP_S$, and its angular, spatial, and temporal distributions are easily computed because they are similar to those of the electron beam.
 
\section{Examples}
To examine the effectiveness of the numerical method proposed in the previous section, we performed FEL simulations using the same parameters as those shown in Fig. 1; in the following discussions, the simulation results without and with the proposed method are refereed to as (A) and (B), respectively. Figure 5 summarizes the gain curves for the 1st, 2nd and 3rd harmonics, with the alphabets and numbers indicating the simulation method and harmonic order. For example, (A-1) shows the result for the fundamental radiation without the proposed numerical method.  

\begin{figure}[htb]
\includegraphics[width=\textwidth]{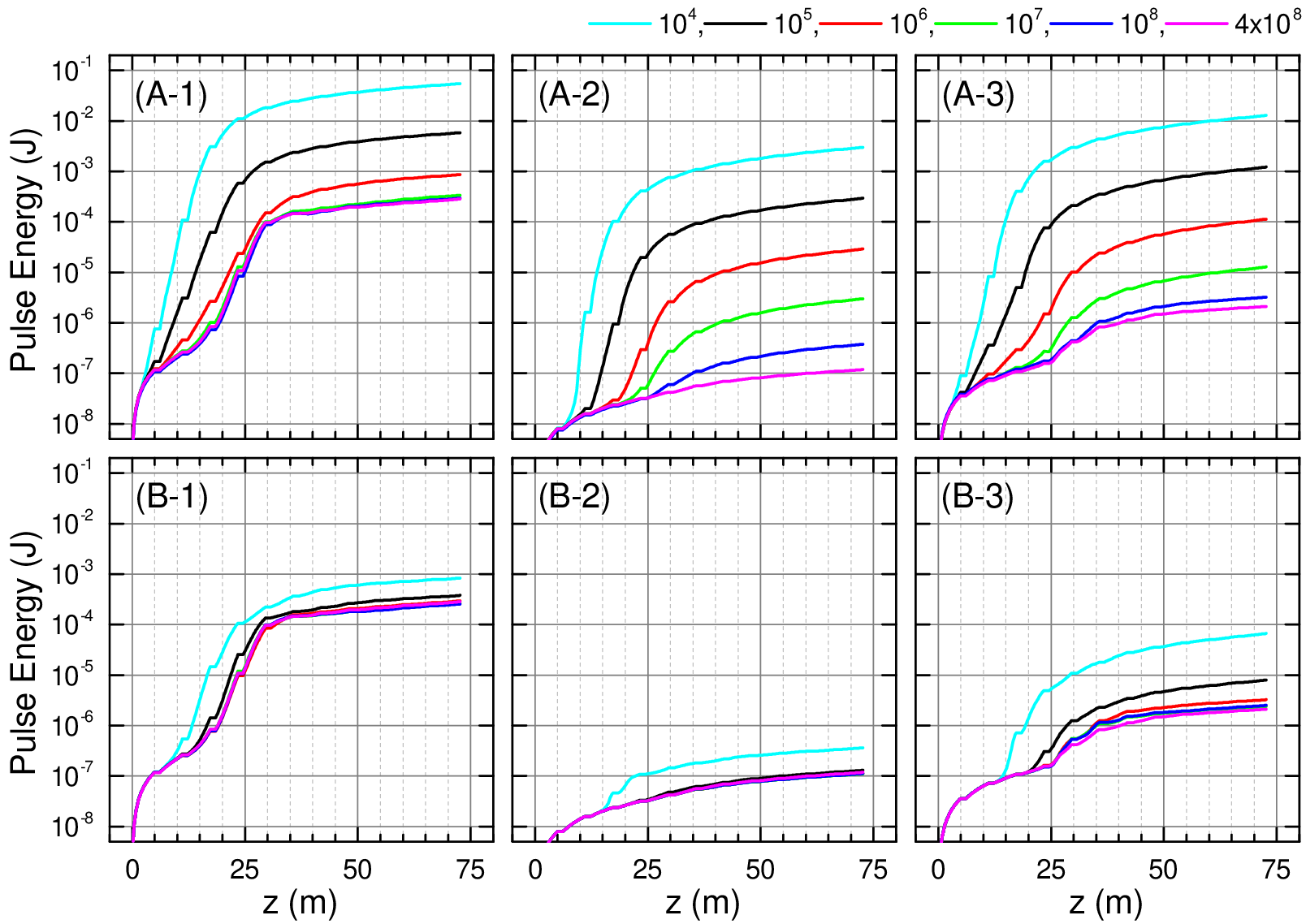}
\caption{Gain curves simulated without (A) and with (B) the proposed numerical method for the (1) 1st , (2) 2nd, (3) 3rd harmonics.}
\end{figure}

Note that two extreme conditions have been added besides those in Fig. 1, which are indicated by cyan and magenta lines; the former shows the results with $M=10^4$ macroparticles, while the latter shows ``rigorous'' simulation results performed without the quiet loading scheme using the real number of electrons ($\sim 4\times 10^8$), and are common to (A) and (B). The gain curves converge to the rigorous results (magenta) when $M$ increases, showing the validity of the simulation with the macroparticle model both in (A) and (B). What should be emphasized more is that (B) reaches the numerical convergence with much fewer macroparticles than (A), meaning the effectiveness of the proposed method.

To examine how the numerical convergence varies as the simulation condition, the fundamental pulse energies at the exit of the undulator ($=\mathscr{E}$) are plotted as a function of $M$ in Fig. 6(a). The dashed line indicates the pulse energy ($=\mathscr{E}_R$) simulated with the real number of electrons, clearly indicating the acceleration of numerical convergence by applying the proposed method.

\begin{figure}[htb]
\includegraphics[width=\figh\textwidth]{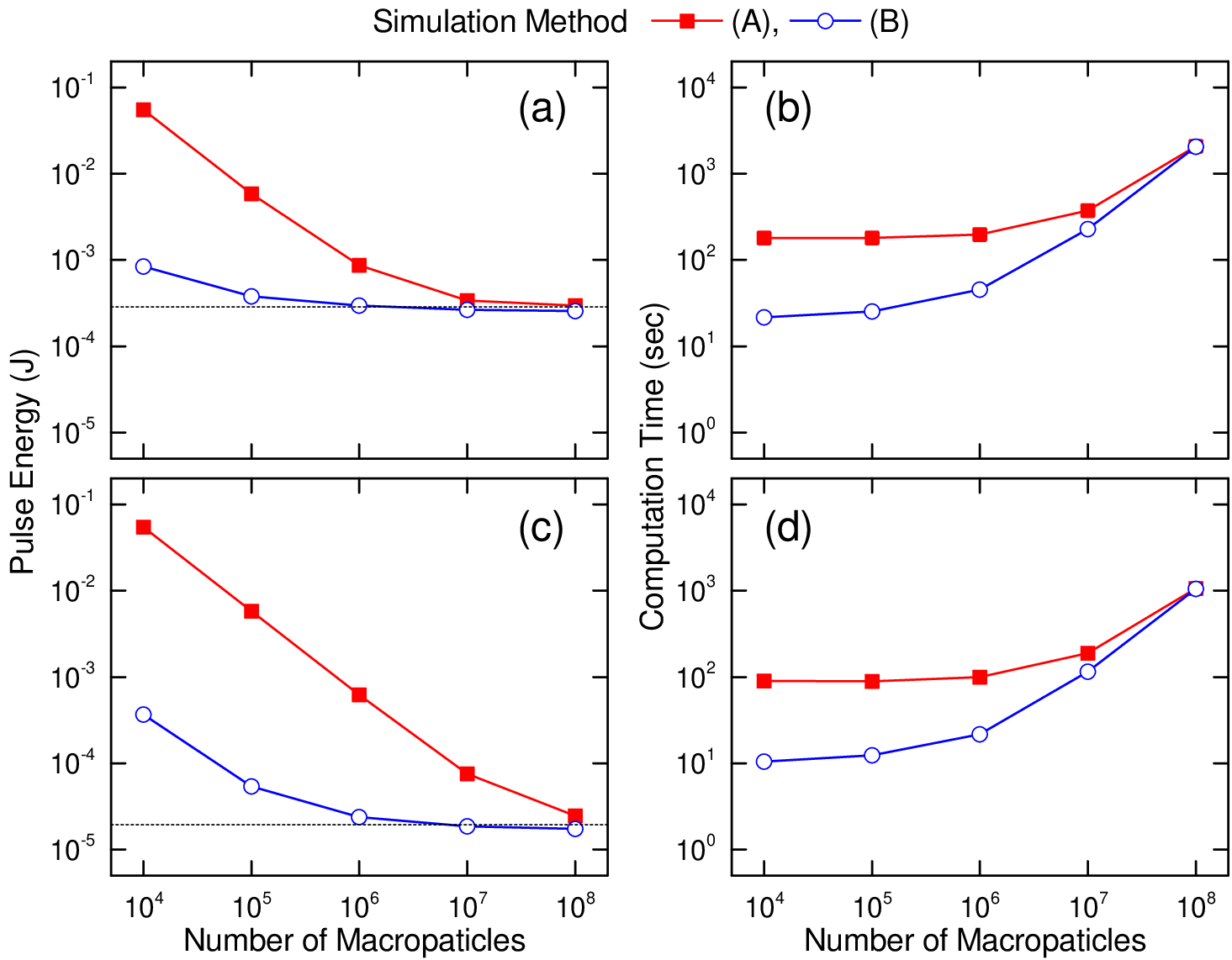}
\caption{Effects of the proposed method to reduce the numerical cost: (a) pulse energy at the undulator exit and (b) computation time, plotted as a function of the number of macroparticles. Results in (c) and (d) are the same as (a) and (b), but with the photon energy of 30 keV.}
\end{figure}

Besides the numerical convergence, the proposed method works to reduce the computation time by enabling the direct computation of radiation field without solving the wave equation; this is clear from Fig. 6(b) where the computation time is plotted as a function of $M$; when $M$ is not large ($M \leq 10^6$) and the dominant numerical process is the computation of radiation field, (B) is much faster than (A), as long as the same number of macroparticles are used. Note that the simulations have been performed with a general computer (Intel Core i9-10850K CPU, 3.60GHz) using 4 MPI (message passing interface) processes, with 3631 slices ($\Delta s=1.49$ nm) and 298 steps ($\Delta z=216$ mm). Although the computation time will become shorter or longer depending on these simulation conditions, it is obvious that the significant reduction by means of the proposed method is universal.

The reduction of the numerical cost by applying the proposed method as found in Figs. 6(a) and 6(b) becomes more significant when the FEL gain is reduced. As an example, Figs. 6(c) and 6(d) summarize the simulation results performed with a lower deflection parameter of 0.5 and the resultant fundamental photon energy of 30 keV. Without the proposed method (A), we need $10^8$ macroparticles or more to reach the numerical convergence, while the simulation with $10^6$ macroparticles gives a reasonable result when the proposed method is applied (B). Note that the number of steps is 3 times fewer than that for 10-keV simulations to have the same number of slices. As a result, the computation time is relatively shorter, however, the overall trend that (B) is much faster than (A) is reproduced.

\begin{figure}[b]
\includegraphics[width=\figh\textwidth]{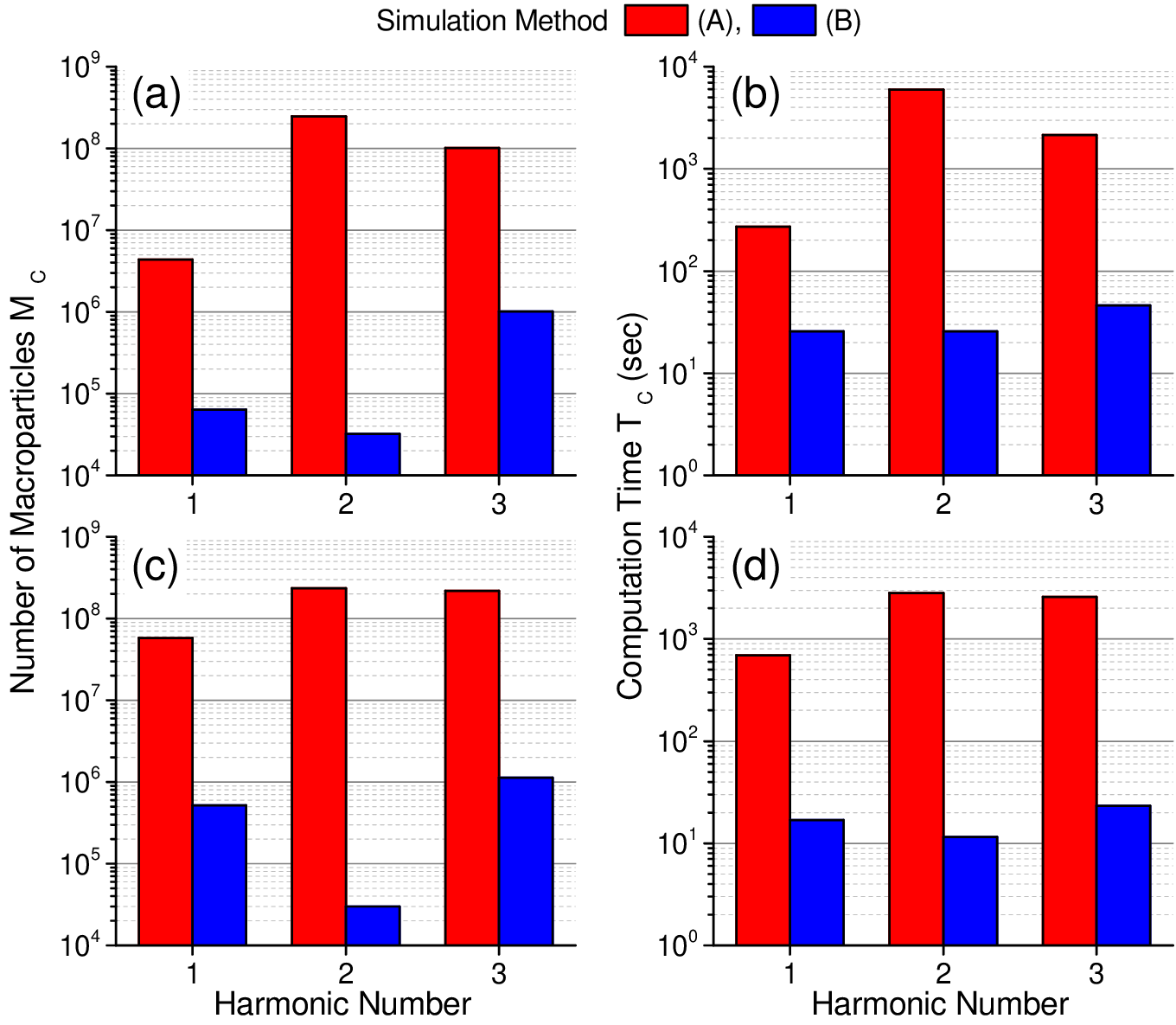}
\caption{Number of macroparticles and computation time to reach numerical convergence.}
\end{figure}

Now let us quantify the actual number of macroparticles to reach the numerical convergence ($M_C$), above which the difference between $\mathscr{E}$ and $\mathscr{E}_R$ is less than 50\%, namely, the condition $\mathscr{E}/\mathscr{E}_r\leq 1.5$ is satisfied. We also define the time to complete a simulation with $M_C$ macroparticles  ($T_C$). Figures 7(a)-7(d) show $M_C$ and $T_C$ determined by interpolating the results shown in Figs. 6(a-d), where we find a significant reduction of the numerical cost by applying the proposed numerical method, and the computation time is 1-2 orders of magnitude shorter, depending on the simulation condition. 

\section{Summary}
We presented a method to accelerate the numerical convergence in FEL simulations. Retrieving the spatially-coherent component reduces the number of macroparticles for convergence, while the direct computation of radiation field simplifies the numerical process. The proposed method significantly reduces the computation time for FEL simulations, which helps the works that may need to repeat a huge number of simulations, such as optimization of the accelerator and undulator configurations to maximize the FEL performance. Note that the numerical method presented in this paper is implemented in {\it SIMPLEX}, which has recently undergone a major upgrade and is available online \cite{URL-simplex-none}.

\bibliography{references}

\end{document}